\documentclass{ws-p8-50x6-00}

\def\bea{\begin{eqnarray}}
\def\beann{\begin{eqnarray*}}
\def\beq{\begin{equation}}
\def\eea{\end{eqnarray}}
\def\eeann{\end{eqnarray*}}
\def\eeq{\end{equation}}

\begin{document}

\title{Heavy ion collisions at intermediate energies in a quark-gluon exchange
framework}
\author{D. T. da Silva and D. Hadjimichef}
\maketitle

\address{ Instituto de F\'{i}sica e Matem\'{a}tica,
Universidade Federal de Pelotas (UFPel) \ \\
CEP 96010-900 Pelotas, R.S., Brazil\\}

\abstracts{
Heavy ion collisions at intermediate energies can be studied in the context
of the Vlasov-Uehling-Uhlenbeck (VUU) model. One of the main features in
this model is the nucleon-nucleon (NN) cross section in the collisional
term. Quark interchange plays a role in the NN interaction and its effect
can be observed in the cross section. We explore the possibility that quark
interchange effects can appear in observables at energies lower than RHIC.
}

\section{Introduction}

In the present one of the most important subjects in hadron and 
particle physics
is the identification of the quark-gluon plasma. It is believed that this
new phase could be observed in high energy heavy ion collisions. In this
regime temperature and density are sufficiently high in order to produce the
deconfinment of the quarks and gluons. At lower energies heavy ion
collisions can be used to probe the conditions related to the nuclear
equation of state. This has been effectively studied in the context of the
Vlasov-Uehling-Uhlenbeck (VUU) model \cite{stocker}.

On the other hand it has been shown that quark interchange effects can play
a role in the nucleon-nucleon interaction at medium energies \cite{dimi,QBD}. 
In this context one can speculate about the appearance of quark interchange
effects in heavy ion collisions at intermediate energies.

In nuclear matter many nucleon quantities become effective ones such as the
radius and/or the cross section. We pretend to parameterize the NN cross
section in the VUU equation by the one obtained from the a quark interchange
mechanism known as the Fock-Tani formalism and investigate the consequences
on the transverse momentum distributions.

\section{The Vlasov-Uehling-Uhlenbeck Model}

The heavy ion collision is described by the classical distribution function
for nuclei. This equation depends on the NN cross section and a potential U
obtained from equation of state. The Vlasov-Uehling-Uhlenbeck (VUU) \cite
{stocker} equation is a differential equation for the classical one-body
phase-space distribution function $f({\bf r},{\bf p} ,t)$ corresponding to
the classical limit of the Wigner function. The VUU equation takes a
familiar form 
\beann
\frac{\partial f}{\partial t}+{\bf v}\cdot \bigtriangledown
_{r}f-\bigtriangledown _{r}U\cdot \bigtriangledown _{p}f &=&-\int \frac{%
d^{3}p_{2}\,d^{3}p_{1}^{\prime }\,d^{3}p_{2}^{\prime }}{(2\pi )^{6}}\;\sigma
\;v_{12}\,\,\delta ^{3}(p+p_{2}-p_{1}^{\prime }-p_{2}^{\prime })
\\
&&\times \left[ f\,\,f_{2}(1-f_{1}^{\,\prime })(1-f_{2}^{\,\,\prime
})-f_{1}^{\,\prime }\,\,f_{2}^{\,\,\prime }(1-f)(1-f_{2})\right]
\eeann
where $f_{2}^{\;\prime }$ is a stands for $f({\bf r},{\bf p}%
_{2}^{\prime },t)$ and correspondingly for the other terms. Using the Skyrme
ansatz (the potential is given by two- and three-body interactions), one
finds a general form for $U=\alpha \rho + \beta\rho^\gamma$ The parameters $%
\alpha,\beta,\gamma$ are fixed\cite{stocker}. The integration of the VUU
equation is done by propagating an ensemble of N events in parallel, with $A$
nucleons simulating the $f$ function. The solution of the transport equation
is not done analytically but in the form of Monte Carlo simulations.

\section{The Fock-Tani Formalism}

The central idea in the Fock-Tani method is the change of representation
concept. The operators of the composite particles are redescribed by ideal
operators which obey canonical anticommutation relations. These ideal
operators act on an enlarged Fock space which is a graded direct product of
the original Fock space and an ``ideal state space''. The ideal operators
correspond to particles with the same quantum numbers as the composite ones
of the system. A change of representation is implemented by means of a
unitary transformation, which transforms the single-composite states into
single-ideal states. When the unitary transformation is applied on the
microscopic quark-quark Hamiltonian one obtains the effective interaction. A
scattering amplitude can be obtained from this procedure and the NN OGEP
(One-Gluon Exchange Potential) is described by an amplitude $h_{fi}$
presented in ref \cite{dimi,QBD}. The Fock-Tani cross-section is given by 
\cite{QBD2}
\beann
\sigma_{NN} =\frac{4\pi ^{5}\,s}{s-4m^{2}}
\int_{-(s-4m^{2})}^{0}\,dt\,|h_{fi}|^{2}
\eeann
where $s$ and $t$ are the Mandelstam variables, $m$ is the nucleon mass.

\section{Results and Discussion}

The cross-section will depend on a parameter $a$ which is the {\sl rms}
radius of the nucleon. In Fig. 1 this can be seen for three values of $a$ in
comparison with the original cross-section of the VUU model. The effect on
the transverse momentum distribution ($p_{x}$) as a function of rapidity 
($y$) is seen in Fig. 2 for a Nb+Nb collision with an impact parameter
$b$. Different from free NN scattering, with quark
interchange, where one has to use $a$ in a range from 0.45 to 0.6 fm 
in order to fit the lower partial waves\cite{dimi2}, in heavy ion collisions 
the ranges from 0.3 to 0.4 fm approach the standard VUU simulations.
Larger radius for the bag, such as 0.7 to 0.8 fm, are far off the
original VUU calculation.

\underline{{\bf Acknowledgements}}

The authors are supported by FAPERGS.

\newpage

\begin{figure}[htb]
\epsfxsize=7cm 
\par
\begin{center}
\epsfbox{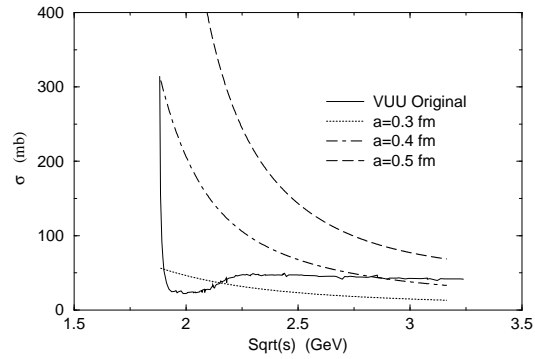}
\end{center}
\caption{The pp cross-section used by the original VUU model and the
modified model with the Fock-Tani cross-section }
\end{figure}


\begin{figure}[htb]
\epsfxsize=6cm 
\par
\begin{center}
\epsfbox{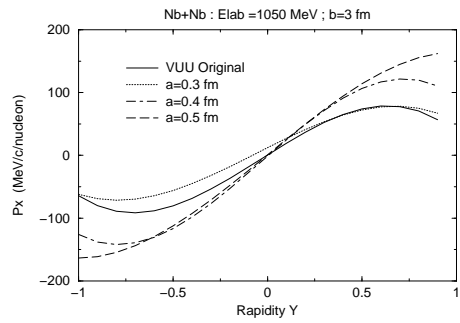}
\end{center}
\caption{Nb+Nb collision: $E_{lab}=1050$ MeV ; $b=3$ fm. }
\end{figure}

\end{document}